\documentclass[useAMS,usenatbib,usegraphicx]{mn2e}

\newcommand{\Msolar}{\mbox{\,$\rm M_{\odot}$}}

\title[NIR integral field spectroscopy of MYSOs]
      {Near-infrared integral field spectroscopy of Massive Young Stellar Objects}
\author[K. Murakawa et al.]{
K. Murakawa$^1$, S. L. Lumsden$^1$, R. D. Oudmaijer$^1$,
B. Davies$^2$,\\ 
{\small \phantom{bliep}} \\
{\LARGE \rm H. E. Wheelwright$^3$, M. G. Hoare$^1$, and J. D. Ilee$^{1,4}$}\\
$^{1}$School of Physics and Astronomy, EC Stoner Building, University of Leeds,
Leeds LS2 9JT\\
$^{2}$Institute of Astronomy, University of Cambridge, Madingley Road,
Cambridge CB3 0HA\\
$^{3}$Max-Planck-Institut f\"{u}r Radioastronomie, Auf dem H\"{u}gel 69,
53121, Bonn, Germany\\
$^{4}$School of Physics and Astronomy, University of St Andrews,
St Andrews KY16 9SS\\
}
\begin{document}

\date{Accepted 1988 December 15. Received 1988 December 14; in original form 1988 October 11}


\maketitle

\label{firstpage}

\begin{abstract}

We present medium resolution ($R\approx5300$) $K$-band integral field
spectroscopy of six massive young stellar objects (MYSOs).  The targets are
selected from the Red MSX Source (RMS) survey, and we used the ALTAIR adaptive
optics assisted Near-Infrared Integral Field Spectrometer (NIFS) mounted on the
Gemini North telescope.  The data show various spectral line features including
Br$\gamma$, CO, H$_2$, and \mbox{He\,{\sc i}}.  The Br$\gamma$ line is detected
in emission in all objects with $v_\mathrm{FWHM}\sim100$ -- 200~kms$^{-1}$.
V645~Cyg shows a high-velocity P-Cygni profile between $-800$~kms$^{-1}$ and
$-300$~kms$^{-1}$.  We performed three-dimensional spectroastrometry to
diagnose the circumstellar environment in the vicinity of the central stars
using the Br$\gamma$ line.  We measured the centroids of the velocity
components with sub-mas precision.  The centroids allow us to discriminate
the blueshifted and redshifted components in a roughly east--west direction in
both IRAS~18151--1208 and S106 in Br$\gamma$.  This lies almost perpendicular to
observed larger scale outflows. We conclude, given the widths of the lines and
the orientation of the spectroastrometric signature, that our results trace a
disc wind in both IRAS~18151--1208 and S106.  The CO $\nu=2-0$ absorption lines
at low $J$ transitions are detected in IRAS~18151--1208 and AFGL~2136.  We
analysed the velocity structure of the neutral gas discs, which we find to have
nearly Keplerian motions.  In IRAS~18151--1208, the absorption centroids of the
blueshifted and redshifted components are separated in a direction of
north-east to south-west, nearly perpendicular to that of the larger scale
$H_2$ jet.  The position-velocity relations of these objects can be reproduced
with central masses of 30~$M_{\sun}$ for IRAS~18151--1208 and 20~$M_{\sun}$ for
AFGL~2136.  We also detect CO $\nu=2-0$ bandhead emission in IRAS~18151--1208,
S106 and V645 Cyg.  The results can be fitted reasonably with a Keplerian
rotation model, with masses of 15, 20 and 20~$M_{\sun}$ respectively.  These
results for a sample of MYSOs can be explained with disc and outflow models and
support the hypothesis of massive star formation via mass accretion through
discs as is the case for lower mass counterparts.
\end{abstract}


\section{Introduction}

\begin{table*}
  \begin{center}
  \caption[]{Physical parameters of the target objects}
  \label{obs_parm}
  \begin{tabular}{llllccccc}
  \hline
  \hline
  \multicolumn{2}{c}{target} & \multicolumn{2}{c}{position$^1$ (J=2000)} &
  $L_{\star}^1$ & $M_{\star}^2$ & spc. type  & $D^1$ & brightness$^1$ \\
  \hline
  & RMS name & \multicolumn{1}{c}{R.A.} & \multicolumn{1}{c}{Dec.} &
  $L_{\sun}$ & $M_{\sun}$ &    & kpc & $m_K$ \\
  \hline
  IRAS 18151--1208  & G018.3412+01.7681  & 18 17 58.1 & --12 07 24.8 &
    2.2$\times10^4$ & 15 & B0$^3$   & 2.8 &  9.3 \\
  AFGL 2136         & G017.6380+00.1566  & 18 22 26.4 & --13 30 12.0 &
    5.3$\times10^4$ &    &               & 2.2 &  7.3 \\
  IRAS 19065+0529   & G039.9018--01.3513 & 19 09 02.5 &  +05 34 42.2 &
    1.0$\times10^4$ &    &               & 3.2 & 11.6 \\
  S106~IRS4         & G076.3829--00.6210 & 20 27 26.8 &  +37 22 47.7 &
    4.0$\times10^4$ & 20 & O9Ve--B0Ve$^4$ & 1.4 &  5.9 \\
  G082.5682+00.4040 & G082.5682+00.4040a  & 20 42 33.7 &  +42 56 51.3 &
    4.2$\times10^3$ &    &               & 1.4 &  9.7 \\
  V645 Cyg          & G094.6028--01.7966 & 21 39 58.3 &  +50 14 20.9 &
    4.3$\times10^4$ & 20 & O7e$^5$       & 4.9 &  6.8 \\
  \hline
  \end{tabular}
  \end{center}
  $^1$ The values are from the RMS data base.
  $^2$ Rough estimation from the spectral types.
  $^3$ \cite{fallscheer11}, and Appendix \ref{discussion}.
  $^4$ \cite{hanson96}.
  $^5$ \cite{cohen77}.
\end{table*}

The formation of high-mass stars ($M_{\star}\ge8M_{\sun}$) is one of the
fundamental issues in astrophysics.  The formation mechanism of low-mass
stars via accretion through a circumstellar disc is generally believed to be
well understood \citep[e.g.][]{shu87}.  However, it was realised relatively
early that the same mechanism for massive star formation potentially suffered
from problems due to the immense radiation pressure from the forming star
\citep[e.g.][]{khan74}.  High-mass stars are usually expected to have short
Kelvin-Helmholtz time scales ($\le10^4$~yr), so that nuclear burning begins
while the gaseous material is still accreting from the natal cloud, and
before the star exceeds $\sim10$\Msolar \citep{shu87}.  The resultant
radiation pressure from that burning acts on both the gas and dust
surrounding the star, counteracting further accretion \citep[e.g.][]{zy07}.
This eventually led to the discussion of models in which massive stars formed
through much more dynamical processes, such as competitive accretion
\citep[e.g.][]{bonnell2004}.  However an alternative mechanism exists which
preserves the connection with low mass star formation.  The high densities
and degree of turbulence and short timescales involved in high mass star
formation regions lead to accretion rates sufficient to ensure the infalling
material can overcome the radiation pressure \citep[e.g.][]{tan2003}.  In
addition, the role of outflows in creating channels for the radiation to
escape, reducing the pressure in the circumstellar environment, had also 
been underestimated \citep[e.g.][]{yorke2002}.

Recent, more detailed, numerical calculations have shown that the conclusions
of McKee \& Tan and Yorke \& Sonnhalter are substantially correct, at least in
showing that massive stars can form in such an environment.  The collimated
outflows do help to reduce the radiation pressure, but also act to generate
further turbulence in the surrounding molecular cloud, effectively helping
sustain the conditions required by the McKee \& Tan model.  Models by Krumholz
et al.\ (2009, 2010), using an approximate treatment of the radiation field,
found that the role of the disc and outflow cavity was crucial.  Kuiper et
al.\ (2010) have shown that a full radiative transfer treatment results in
there being essentially no limit to the mass of star that can be formed, since
the disc effectively self-shields against the star's radiation pressure. More
recently Kuiper \& Yorke (2013) have shown that gas, rather than dust, opacity
above the disc creates an additional shield, so that the ``radiation pressure''
problem is reduced even further.

%


Direct detection of small scale discs around young massive protostars is however
observationally challenging \citep[e.g.][]{kraus10}.  We have pioneered the use
of spatially resolved near-infrared spectroscopy in this field to help address
the issue of the properties of discs around massive protostars, as well as any
small scale outflows present \citep[e.g.][]{davies10}, and this paper 
follows on from our earlier work.

The near-infrared (NIR) spectra of massive young stellar objects (MYSOs) show a
variety of important diagnostic emission and absorption features.  For example,
hydrogen recombination lines such as Br$\gamma$ (\mbox{H\,{\sc i}} $n=7-4$ at
2.166~$\mu$m) are emitted from ionized regions.  They are often used as tracers
of the mass accretion in lower mass stars
\citep[e.g.][]{gl96,najita96,muzerolle98,fe01}, as well as of stellar and disc
winds \citep[e.g.][]{persson84,bhd95,nisini95,cr98,drew98,ishii01} at all masses.  The former show low velocity ($|v|\la100$ -- 200~kms$^{-1}$)
emission, often with associated weak red absorption features from the infalling
gas, while the latter show wider velocity ranges ($v\la200-600$~kms$^{-1}$) and
sometimes a P-Cygni absorption profile suggesting an outflowing wind
\citep[e.g.][]{drew98,kurosawa11,lumsden12}.  Another interesting feature is
the first overtone ro-vibrational spectrum of carbon monoxide around
2.3~$\mu$m.  The CO bandhead is often seen in emission and is thought to
arise from accretion discs of neutral gas with temperatures ranging between
$\sim$2000~K and $\sim$5000~K.  These features are therefore useful in tracing
structures ranging from the hot ionized gas to a warm dusty disc.
If the gaseous CO discs show Keplerian rotation at a radius of 0.1~AU, and
the mass of the central star is $\sim10~M_{\sun}$, the rotation velocity is
$\sim$300~kms$^{-1}$, which can be analysed by medium and high resolution
spectroscopy.  In fact, the observed CO $\nu=2-0$ bandhead emission in young
stellar objects as a whole can be explained with Keplerian rotating disc models
\citep[e.g.][]{bw84,carr89,chandler95,najita96,martin97,kraus00,bt04,wheelwright10}.

Recently, some results of NIR integral field spectroscopy using 8~m class
telescopes have been reported \citep{davies10,goto12,stecklum12}.  With this
method, we obtain the images and spectra simultaneously and can perform
spectro-astrometry.  The technique allows us to measure the position centroids
of the circumstellar features with sub-mas accuracy.  Using 3D
spectro-astrometry, \citet[][ hereafter D10]{davies10} showed that the
Br$\gamma$ emission line in the MYSO W33A reveals evidence for a fast bipolar
jet ($v\sim600$~kms$^{-1}$) on sub-milliarcsec scales (cf.\,the intrinsic
angular resolution of an 8m telescope of $\sim$100~mas in the $K$ band).  From
a velocity analysis of the CO $\nu=2-0$ bandhead, they estimated the stellar
mass to be 10 -- 15~$M_{\sun}$.  We have extended the same technique as
used by D10 to a sample of MYSOs selected from the RMS \citep[Red MSX Source
survey,][]{lumsden02} and search for the aforementioned signatures.
Sect.\,\ref{obs}, describes the observations and the data reduction procedure,
and presents the images and spectra.  We perform the analysis of the Br$\gamma$
emission lines and the velocity structures from the CO spectra of selected
targets in Sect.\,\ref{spcline}.  Appendix \ref{discussion} provides a
discussion of individual objects.

\begin{table}
  \begin{center}
  \caption[]{NIFS integral field spectroscopy observations}
  \label{obs_parm}
  \begin{tabular}{lrccc}
  \hline
  \hline
  target    & int. time & \multicolumn{2}{c}{standard} & date \\
  \hline
            & \multicolumn{1}{c}{sec}   &   ID    & spc. type \\
  \hline
  I18151    &   2880    &  HR 6956  &    A5IV   & 2011-07-17 \\
  AFGL~2136 &    900    &  HR 7288  &    A3V    & 2011-07-15 \\
  I19065    &   1440    &  HR 8291  &    A2V    & 2011-07-15 \\
  S106      &    510    &  HR 7734  &    A0V    & 2011-07-14 \\
  G082      &   1080    &  HR 7752  &    A1V    & 2011-07-13 \\
  V645~Cyg  &    200    &  HR 8291  &    A2V    & 2011-07-19 \\
  \hline
  \end{tabular}
  \end{center}
\end{table}

\section{NIFS Integrated field spectroscopy}
\subsection{Observations}\label{obs}

We observed six MYSOs using the Gemini Near-Infrared Integral Field
Spectrograph \citep[NIFS; ][]{mcgregor03} on the nights between 2011 July 13
and July 19.  The target objects are IRAS~18151--1208 (hereafter I18151),
AFGL~2136, IRAS~19065+0529 (hereafter I19065), S106~IRS4 (hereafter S106),
G082.5682+00.4040 (hereafter G082), and V645~Cyg, which were selected from
  the RMS catalogue \citep[][]{lumsden13}.  These, together with W33A from
  BD10, form a complete set of objects for which the central star was
  sufficiently bright for infrared spectroscopy, but also, more crucially,
  where a bright optical guide star was present within 20 arcseconds of the
  target.  They are a rather heterogeneous collection of objects in other
  regards.  I19065 is listed as a weak compact HII region in the RMS catalogue,
  S106 has elements of both a massive protostar and an HII region. In both
  cases we centred the NIFS aperture on the apparent exciting star.  The basic
  properties of all sources (distance, luminosity, apparent $K$ band magnitude)
are listed in Table \ref{obs_parm}.  A summary of the known properties of these
objects and how these tie in with the current observations can be found in the Appendix.

The instrument acquires a $3\times3$~arcsec$^2$ field of view
which is separated into 29 slices and is recorded on a two-dimensional infrared
detector with 2048$\times$2048 pixels.  The $K$-band spectrum with a wavelength
range between 2.0 and 2.4~$\mu$m is obtained, with a spectral resolution of
$R\approx5500$ corresponding to a velocity resolution of $\approx55$kms$^{-1}$.
The ALTAIR adaptive optics system was used to correct the wavefront of the
target object for all sources except S106, for which no suitable guide star was
available.  The system uses a natural-guide star for tip-tilt correction and
the laser-guide star for wavefront sensing.

In our observations, the target objects and the sky field were observed in
turns.   The offset of the sky field is between 20~arcsec and 30~arcsec.
In addition to the science targets, we observed standard stars to correct for
telluric absorption lines.  The Gemini facility calibration plan was used.
This includes the dark frames, the flat field data, Argon-Xenon (Ar-Xe) arc
lamp for wavelength calibration, and Ronchi screen images to calibrate the
spatial distortion of the NIFS data.  The observing log is provided in
Table \ref{obs_parm}.

\subsection{Data reduction}\label{reduction}

Dark frames and flat field frames were constructed for each observation date.
High ADU counts in the dark frames and low counts in the flat field frames are
assumed to be due to hot and dead pixels, respectively.  Bad pixel frames that
include this information were created.

The preprocessing, the dark frame subtraction, flat fielding, and removal of
bad pixels were performed for the targets, standards, arc lamp, and Ronchi
frames.  In the next step, the Ar-Xe lamp data were used for wavelength
calibration, the Ronchi screen data were used to correct the image distortion,
and the two-dimensional images were converted into three-dimensional
(x,y,$\lambda$) cubes.  In the data before this conversion, the pixel scales
are 0.043~arcsec in the space axis in each slice and 0.103~arcsec of the
position increment of the neighbouring slice.  Each wavelength element (2
pixels) has $\Delta\lambda\approx$0.00041~$\mu$m which equates to a velocity
width of $\approx$56~kms$^{-1}$.

The sky frames were then subtracted from the target frames and the
correction for the telluric absorption features was made.  The telluric
standard stars have spectral types around A0V, in which the metal
lines are relatively weak and infrequent \citep{vacca03}.  The
standard stars show a strong, intrinsic Br$\gamma$ absorption line, which 
needs to be removed from the standard star data before they can be used to
correct for the telluric absorption in the targets.
For this purpose, the Br$\gamma$
absorption line was fitted using the \textsf{de-blend} command of the
\textsf{onedspec} task of the IRAF data reduction software and the modeled
function was removed from the standard star spectrum.  The target data
were then divided by the corrected standard star data and were multiplied with
a blackbody function with an effective temperature corresponding to their
spectral type. 

\begin{figure*}
  \centering
  \resizebox{\hsize}{!}{\includegraphics{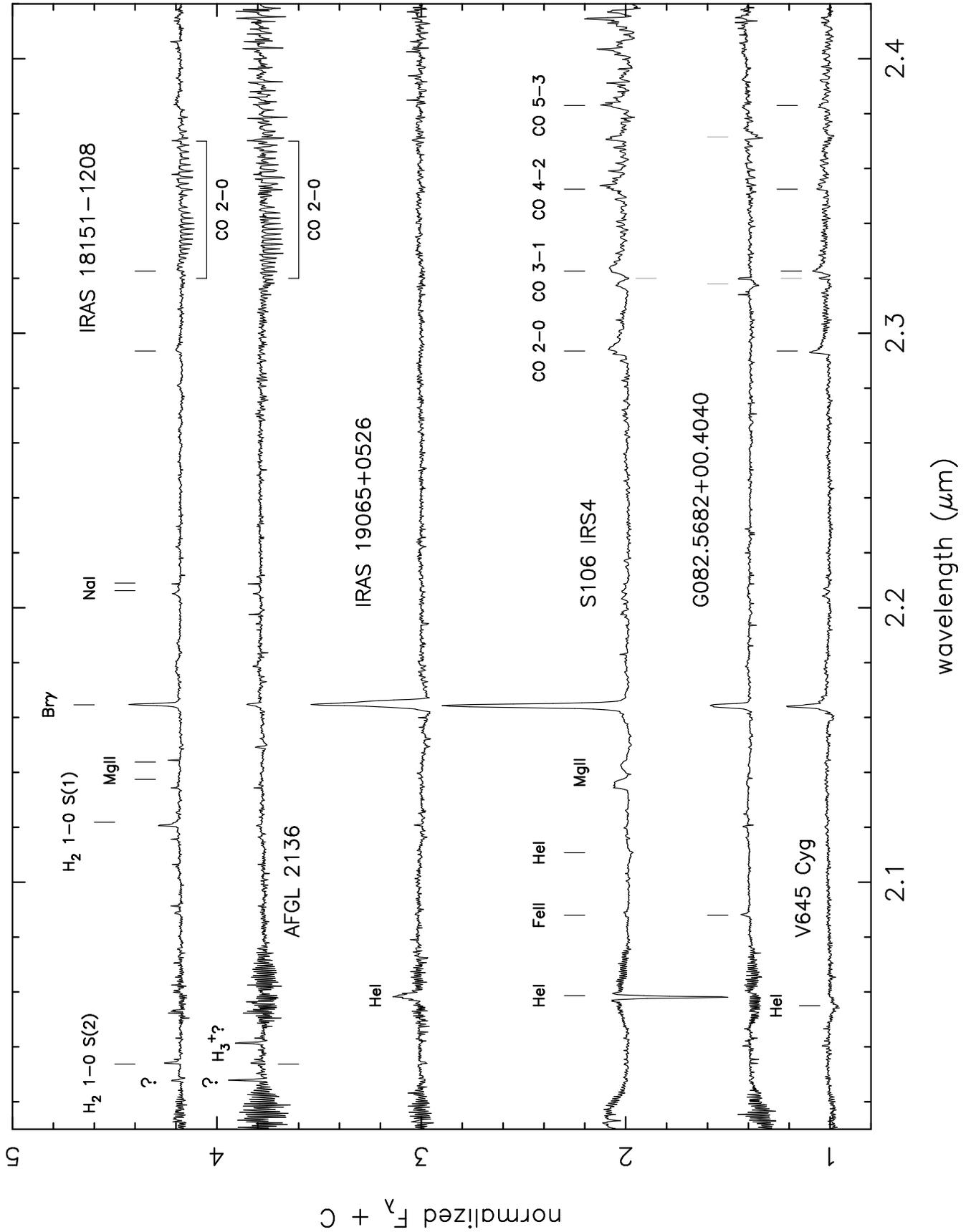}}
  \caption{$K$-band spectra of our target objects obtained using the
           Gemini-north$/$NIFS instrument.
         }
  \label{spectra}
\end{figure*}

\begin{table*}
  \begin{center}
  \caption[]{Identified emission and absorption lines in our observations}
  \label{detline}
  \begin{tabular}{lcll}
  \hline
  \hline
  \multicolumn{1}{c}{target} & Br$\gamma$ & CO & others \\
  \hline
            & $v_\mathrm{LSR}^1/v_\mathrm{FWHM}/v_\mathrm{FWZI}$ & 2.3~$\mu$m$^3$ & \\
  \hline
  I18151    &  $34.2(32.8)/108/\pm158$ &  $\nu=2-0^4$, $3-1$ & H$_2$ 1--0 S(1), S(2), \mbox{Mg\,{\sc ii}}, \mbox{Na\,{\sc i}}\\
  AFGL~2136 &  $21.4(22.5)/133/\pm114$ &  $\nu=2-0^4$ & H$_2$ 1--0 S(1), S(2), \mbox{Na\,{\sc i}} \\
  I19065    &  $18.0(12.5)/211/\pm258$ &  no & \mbox{He\,{\sc i}} \\
  S106      &  $8.09(-1.7)/190/\pm220$  & $\nu=2-0$, $3-1$, $4-2$, $5-3$ & \mbox{He\,{\sc i}}, \mbox{Fe\,{\sc ii}} \\
  G082      & $-2.87(-4.1)/215/\pm230$ &  no & \mbox{Fe\,{\sc ii}} \\
  V645~Cyg  & $-44.4(-43.9)/157/+396^2$ &  $\nu=2-0$, $3-1$, $4-2$, $5-3$ & \mbox{He\,{\sc i}} \\
  \hline
  \end{tabular}
  \end{center}
  $^1$ The values in the parenthesis are from the RMS data base.
  $^2$ Since the feature shows a P~Cyg profile, the value is provided only for
       the redder side.
  $^3$ Vacuum wavelengths of the bandhead are 2.29353~$\mu$m for $\nu=2-0$,
  2.32265~$\mu$m for $\nu=3-1$, 2.35246~$\mu$m for $\nu=4-2$ and 2.38295~$\mu$m
  for $\nu=5-3$ \citep{tokunaga00}.
  $^4$ Absorption in the $R$ and $P$ branches is also detected.
\end{table*}

\begin{figure*}
  \centering
  \resizebox{\hsize}{!}{\includegraphics{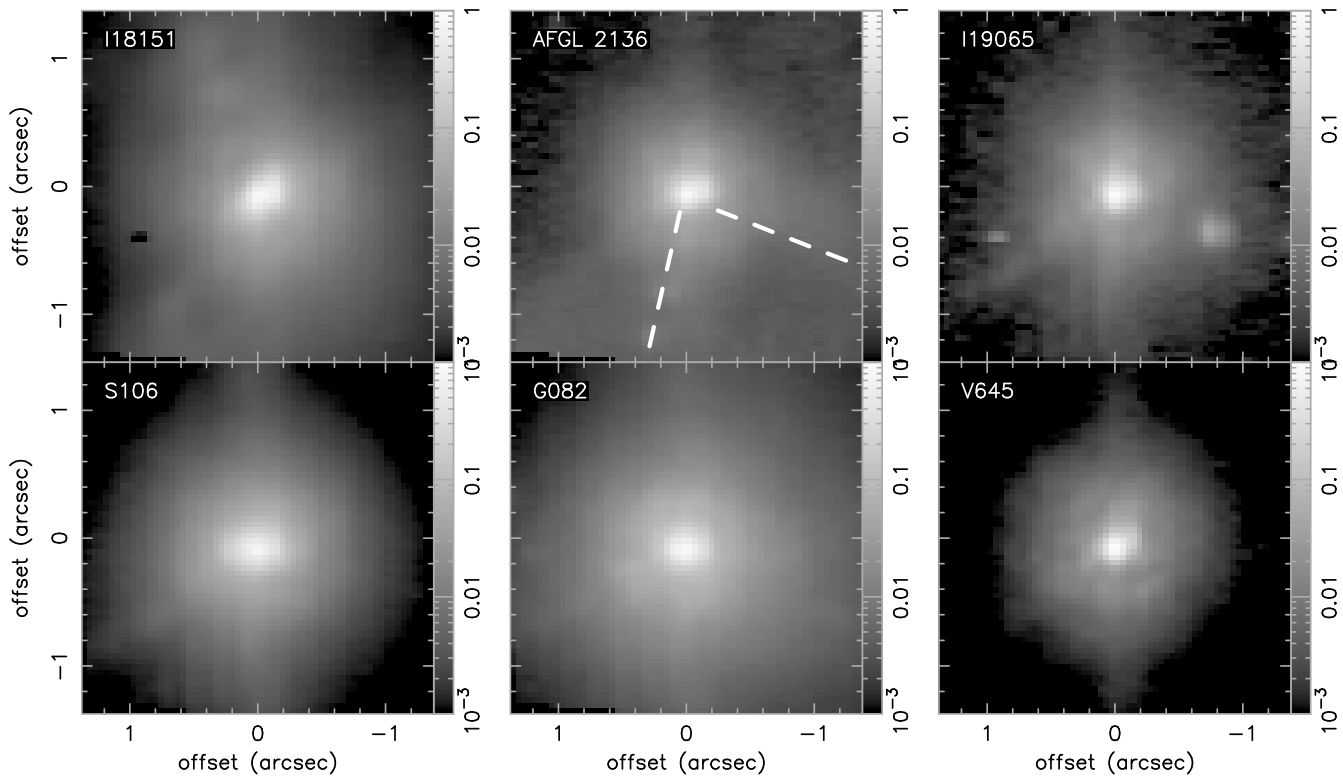}}
  \caption{$K$-band continuum images of our target objects.  North is up and
    East to the left, except for V645~Cyg, where the image is rotated
    counterclockwise by 20\degr.  The surface brightness is normalised by the
    maximum value.
    In AFGL~2136, the \textsf{V}-shaped feature is indicated with dashed line.
    }
  \label{images}
\end{figure*}

The angular resolution was determined by measuring the full width at half
maximum of the telluric standards, which are single stars.  The typical values
are $\sim$0.1~arcsec, which are constant through the entire wavelength range.

For the one-dimensional spectra, where all the fluxes at individual wavelength
channels are integrated through the $(x,y)$ space, the typical SNRs are found
to be 100 -- 300 per wavelength channel.  The spectral resolution is measured
from unresolved telluric absorption lines in the telluric standard stars.
The typical values are $\sim$0.00037~$\mu$m ($\Delta v\sim48$~kms$^{-1}$).
We also measured the statistical variations of the wavelengths of the
telluric absorption features of the standard stars.  By comparing with
the wavelength channel positions of these features of stars that were
observed at different times and dates, we obtained an error of
$\sim$$10^{-5}~\mu$m ($\sim$$1.3$~kms$^{-1}$), which is about one tenth of
one wavelength channel.

I19065 and V645~Cyg were also observed on 2011 July 17 and 15, respectively, in
addition to the dates that are indicated in Table~\ref{obs_parm}.  The
integration times of the targets are 1980 sec. for I19065 and 700 sec. for
V645~Cyg.  For I19065, the SNR of the spectrum is $\sim$100, which is worse
than that of the data taken on July 15 ($\sim$140).  This is due to
non-photometric conditions at the time of observing I19065 on July 17.  This
leads to very poor subtraction of the night sky emission lines in the object
minus sky pairs.  For V645~Cyg, the detector output signal level was saturated,
and hence not suitable for spectroastrometry.  For the observations of July 19,
the exposure time was reduced from 25~seconds to 20~seconds,
  and a neutral density
filter \textsf{KG3\_ND\_FILTER\_G5619} with a transparency of 5\% was inserted
to avoid the saturation.  The signal-to-noise in the nebula in the data using
the filter are lower, but inspection of the unsaturated parts of the image from
July 15 shows no additional information is present.  We therefore prefer to use
only the data from July 19 in this paper for V645~Cyg, and only that from July
15 for I19065.

\section{Results}\label{results}

Here we describe the spectral lines and images detected in our data, 
we defer detailed analysis of interesting lines until
Sect.\,\ref{spcline}.

\subsection{2.0 to 2.4~$\mu$m spectra}

Figure\,\ref{spectra} presents the 2.0 to 2.4~$\mu$m spectra of the target
objects.  To highlight the emission and absorption lines, the spectra are
normalised by a fourth-degree polynomial fit to the continuum.
Table\,\ref{detline} lists the lines identified in our observations.

In all target objects, the Br$\gamma$ line is detected in emission.  The
Br$\gamma$ emission of V645~Cyg shows a P-Cygni profile. I19065 may show
  weak photospheric absorption.  A series of CO emission and absorption
features is detected between 2.3 and 2.4~$\mu$m.  In I18151 and AFGL~2136, the
$R$ and $P$ branches of $\nu=2-0$ lines at low $J$ transitions are in
absorption.  Bandhead emissions of $\nu=2-0$, $3-1$, $4-2$, and $5-3$ are
detected in S106 and V645~Cyg.  In I18151, $\nu=2-0$, $3-1$ and $5-3$ are
seen -- any $4-2$ emission is masked by the CO absorption.  In S106,
G082, and V645~Cyg, absorption features are detected around 2.32~$\mu$m and
2.372~$\mu$m, which are indicated with grey lines in Fig.\,\ref{spectra}.
These features vary significantly when other telluric standard spectra are
used.  Hence, we conclude that these are residual telluric features that were
not fully corrected for.  Other lines that are identified are the molecular
hydrogen lines in I18151 ($\nu=1-0$ S(1) and S(2)).  For \mbox{He\,{\sc i}},
absorption (S106 and V645~Cyg) and emission (I19065) lines at 2.05869~$\mu$m
($2^1P$--$2^1S$) and an absorption line at 2.1127~$\mu$m ($4^{1,3}S$--$^{1,3}P$
in S106) are detected.  We also identified \mbox{Mg\,{\sc ii}} in S106,
\mbox{Na\,{\sc i}} in I18151 and AFGL~2136, and \mbox{Fe\,{\sc ii}} in S106 and
G082.  In AFGL~2136, an emission line is seen at 2.041~$\mu$m.  The
  2$\nu_2 (\nu_2=2-0)$ Q(3--0) transition of H$_3^+$ falls at this wavelength
but the lack of other transitions in the same range from the overtone
  suggests this is an unlikely identification.

\subsection{NIFS images}

\begin{figure}
  \centering
  \includegraphics[width=8.5cm]{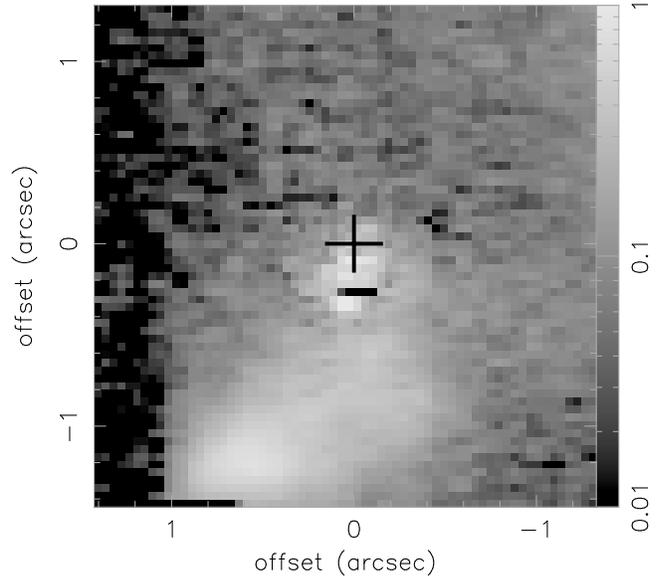}
  \caption{H$_2$ $\nu=1-0$ S(1) image of I18151.  The cross at the zero-point
           indicates the position of the flux peak of the $K$-band continuum.
         }
  \label{h2_i18151}
\end{figure}

Figure\,\ref{images} shows the $K$-band continuum images.  I18151 exhibits a
knot at NNE and a lobe at a distance of $(0.63, -1.08)$ in arcsec from the
central star IRS~1. When comparing with the images published by \cite[][
  hereafter D04]{davis04}, at first glance our image looks rotated by
$180\degr$, e.g.\,a lobe is seen at east in their image in their Fig.\,6a.
However, this is not the case.  The ring-like feature in their image is at a
distance of about 2~arcsec from the central star, while the feature in our
image is at $\sim$1~arcsec.  Figure\,\ref{h2_i18151} shows the H$_2$ $\nu=1-0$
S(1) image.  We see a bright lobe at south-east, which corresponds to the lobe
in our $K$-band continuum image.  This feature is actually the SE jet
identified by D04 (see their Fig.\,6c).  We should note that our AO assisted
data have a much higher resolution than D04's images which were obtained using
a 4~m telescope under the seeing limit (although the seeing or beam size is not
described in their paper).  As a result our images show much more detailed
structures.  In particular, the jet is pointing south nearer to the star.

The AFGL~2136 image reveals a \textsf{V}-shaped feature expanding toward the
south-west.  In the previous $K$-band image obtained using the 8~m Subaru
telescope \citep{murakawa08}, the central star (IRS~1) feature looks nearly
spherical, however their observations were not performed with AO due to the
lack of a a bright reference, while here we use laser guide stars and as
a result achieve a higher resolution.  The \textsf{V}-shaped feature is
a little puzzling.  The southern arm corresponds to the S-lobe(s)
\citep{kastner92,murakawa08}.  However, the western arm was not detected in
the previous observation.  Although an extended filament feature is detected
at south-east in the previous images, identified as SE-lobe \citep{kastner92},
this is detached from IRS~1.  The CO $J=2-1$ emission line image shows a pair
of blueshifted (14 to 20~kms$^{-1}$) and redshifted (24 to 30~kms$^{-1}$)
components, which extend toward south-east and north-west, respectively
\citep{kastner94}.  The NIR polarimetric images clearly show a centro-symmetric
vector pattern surrounding  IRS~1 and a polarization disc extending at
a position angle of $\sim$45$\degr$ \citep{minchin91,kastner92,murakawa08}.
It is reasonable to assume that  IRS~1 is the central (proto-)star powering
the CO outflow and illuminates the surrounding lobes, which is probably
the cavity wall, and the polarization disc is a signature of the presence of
an optically thick dust disc$/$torus.  Although it is not conclusive, taking
into account these geometrical configurations as well as a detection of a lobe
at NNW, the western arm revealed in our image could be the counterpart of the
NNW lobe and these two features are the cavity wall of the northern nebulosity.

The other images are point source like.  The measured FWHM of the images are
$\sim$0.2~arcsec apart from V645~Cyg for which it is 0.12~arcsec.
S106~IRS4 is a deeply embedded O-type young stellar object that excites
the \mbox{H\,{\sc ii}} region S106
\citep{sharpless59,sibille75,gehrz82,richer93}.  The previous near-infrared
imaging of an arcmin scale field of view have shown a butterfly shaped
nebulosity \citep[e.g.][]{saito09}.  The nebulosity is split by an optically
thick dark lane \citep[$A_V\sim21$~mag,][]{felli84}.  It would be reasonable
to expect that our image with a small field of view shows the central star
feature seen through the dust disc and the scattered light, which causes
a slightly extended nebulosity (cf.\,the PSF FWHM size of the telluric standard
star is about 0.1~arcsec).  The V645~Cyg star forming region has three optical
knots and our target is N0 \citep{cohen77}.  The previous near-infrared images
detected an arc-like feature around the central star
\citep[e.g.][]{cohen77,goodrich86,minchin91,clarke06}.  Because of its
characteristic appearance, the nebula is named the ``duck nebula''
\citep{goodrich86}.  With improved spatial resolution observations,
the appearance does not look like a duck anymore, but the name has stuck.
The surface brightness of the central star is significantly higher than that
of the arc-like feature.  This and the FWHM, which is close to that of the
standard stars, suggests that the nebulosity is optically thin in the optical
and near-infrared.

\begin{figure}
  \centering
  \includegraphics[width=8.5cm]{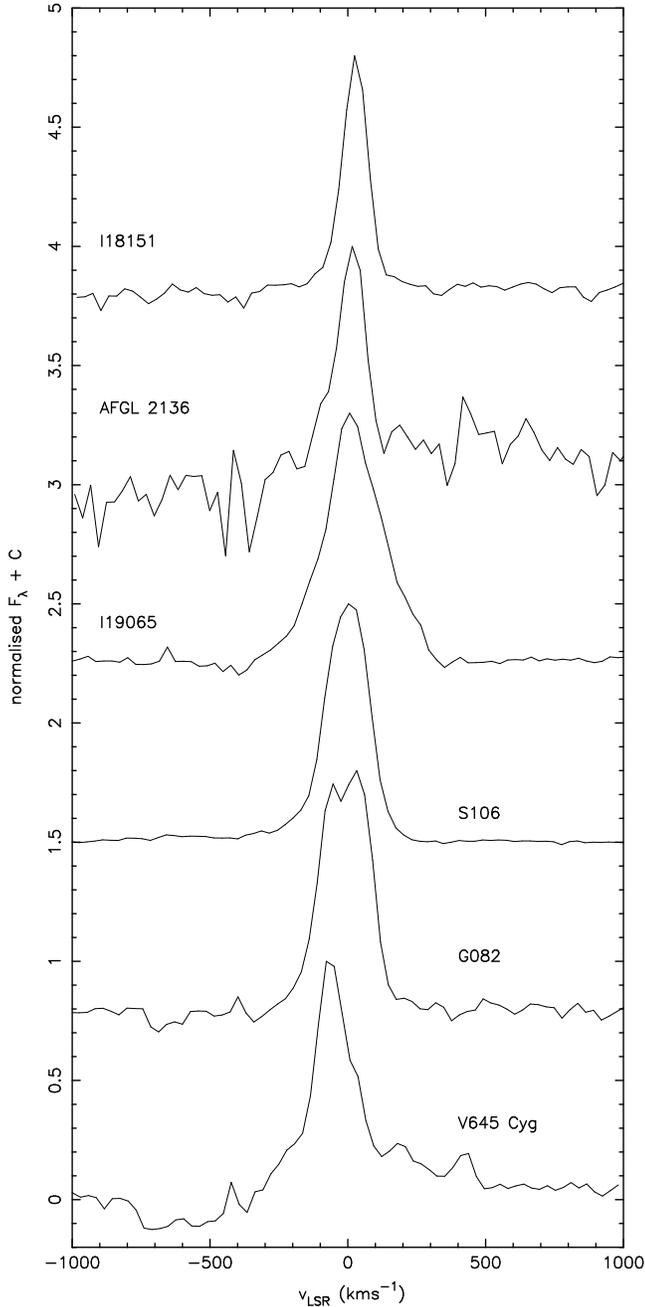}
  \caption{The Br$\gamma$ spectral feature on velocity scale.
         }
  \label{BG_vel}
\end{figure}

\begin{figure*}
  \centering
  \resizebox{\hsize}{!}{\includegraphics{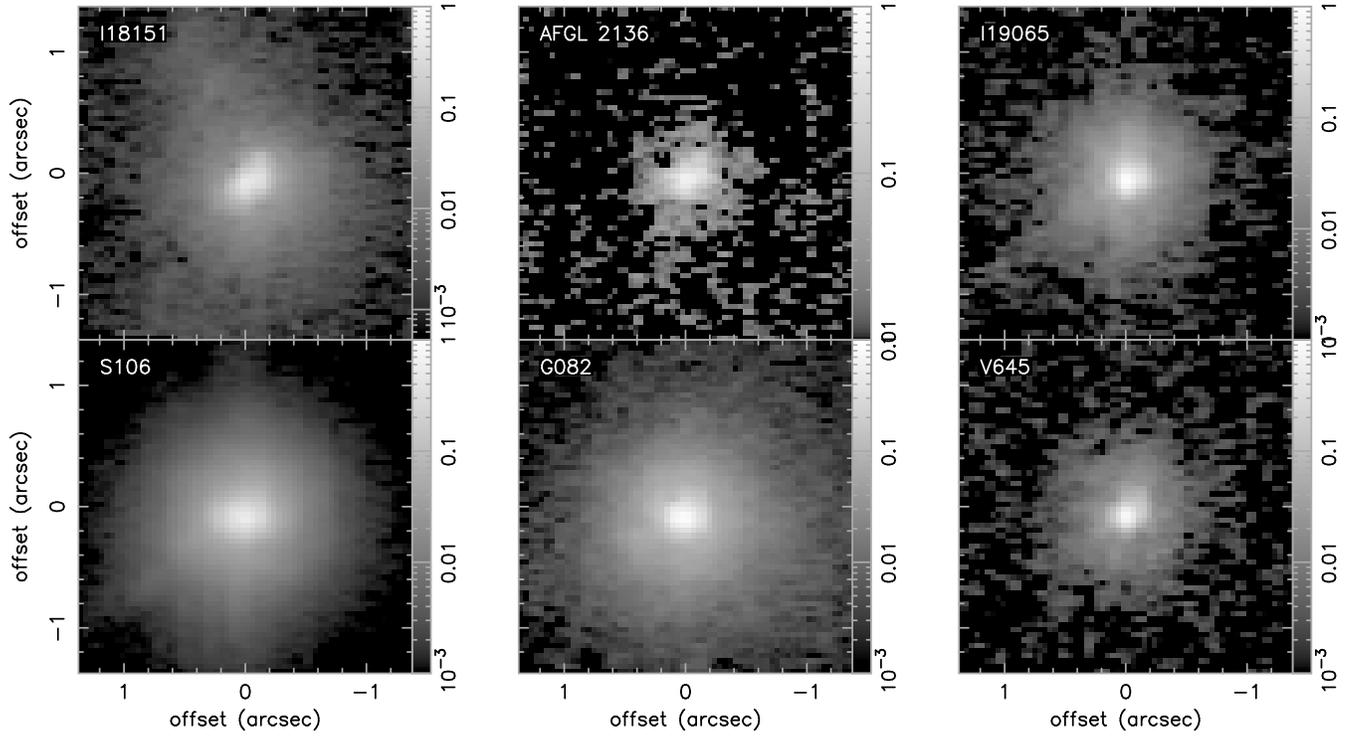}}
  \caption{The Br$\gamma$ images.  The orientation of the images are the same
           as Fig.\,\ref{images}.
         }
  \label{imgbg}
\end{figure*}

\begin{figure*}
  \centering
  \includegraphics[width=17.0cm]{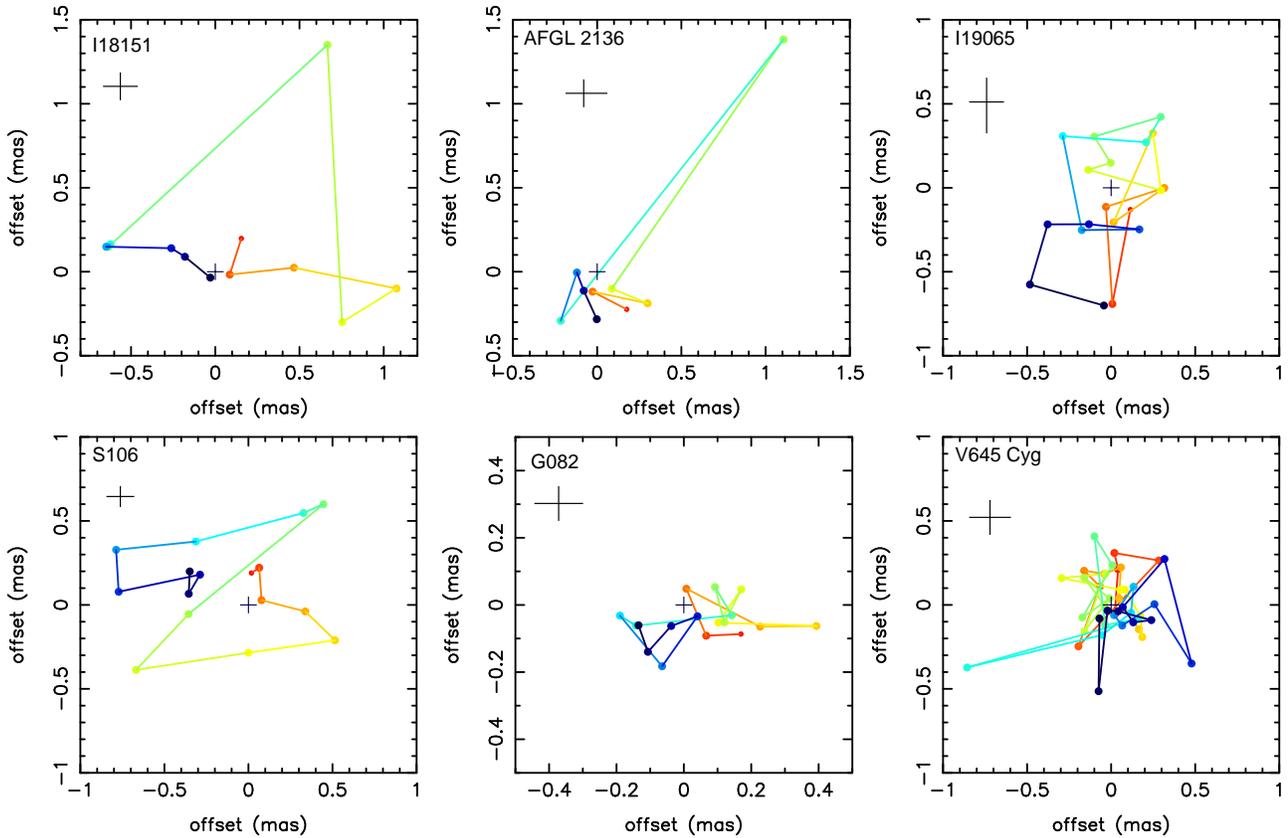}
  \caption{Spectro-astrometry across the Br$\gamma$ emission line.
           The coloured dots indicate the measured centroid positions at
           individual velocity channels.  The velocity ranges are between
           $-800$~kms$^{-1}$ (indicated with blue) and +400~kms$^{-1}$
           (indicated with red) for V645~Cyg and between $\pm v_\mathrm{FWZI}$
           for other objects.  The zero-point indicates the position centroid
           of the nearby continuum.
         }
  \label{BG_spcast}
\end{figure*}

Several spectral lines are detected in the one-dimensional spectra, as
presented in the previous section.  The images of these lines look similar to
those of the continuum apart from the molecular hydrogen image of I18151.
We will constrain our further discussion to the Br$\gamma$ line
(Sect.\,\ref{brg}), the CO $\nu=2-0$ emission and absorption (Sect.\,\ref{co}),
and the \mbox{He{\sc i}} lines (Sect.\,\ref{helium}).

\section{Spectral line data}\label{spcline}
\subsection{Br$\gamma$ emission}\label{brg}

Figure \ref{BG_vel} presents the Br$\gamma$ spectra on a $v_{LSR}$ scale.  The
velocities with respect to the local standard of rest, $v_\mathrm{LSR}$, the
full width at half maximum $v_\mathrm{FWHM}$, and the full width at zero
intensity $v_\mathrm{FWZI}$ are listed in Table\,\ref{detline}.  The
$v_\mathrm{LSR}$ values in the parenthesis are from the RMS sources.  The
discrepancies between the RMS measurements and ours can be explained with the
uncertainties of the measured velocity (see Sect.\,\ref{reduction}) with the
exception of I19065 and S106.  The deviation seen in S106 is relatively small,
and the value derived for Br$\gamma$ is consistent with that seen in
\citet{lumsden12}.  In I19065, CO $J=1-0$ emission is detected at two
  velocities \citep{urquhart07}, with  $v_\mathrm{LSR}$ of +12.5~kms$^{-1}$
  and +48.4~kms$^{-1}$. The latter of these is coincident with NH$_3$ (1,1)
  emission, as well as a water maser complex.  It is possible that the redder
  component is from another cloud which is located within the aperture
  ($\sim$30~arcsec) of the molecular line observations but not within our
  observations.  No other equally bright mid-infrared sources are evident
  within such an aperture however as seen from the RMS database.  The correct
  molecular line counterpart to the infrared source presented here cannot be
  resolved with the data currently available however.

\begin{figure}
  \centering
  \includegraphics[width=8.5cm]{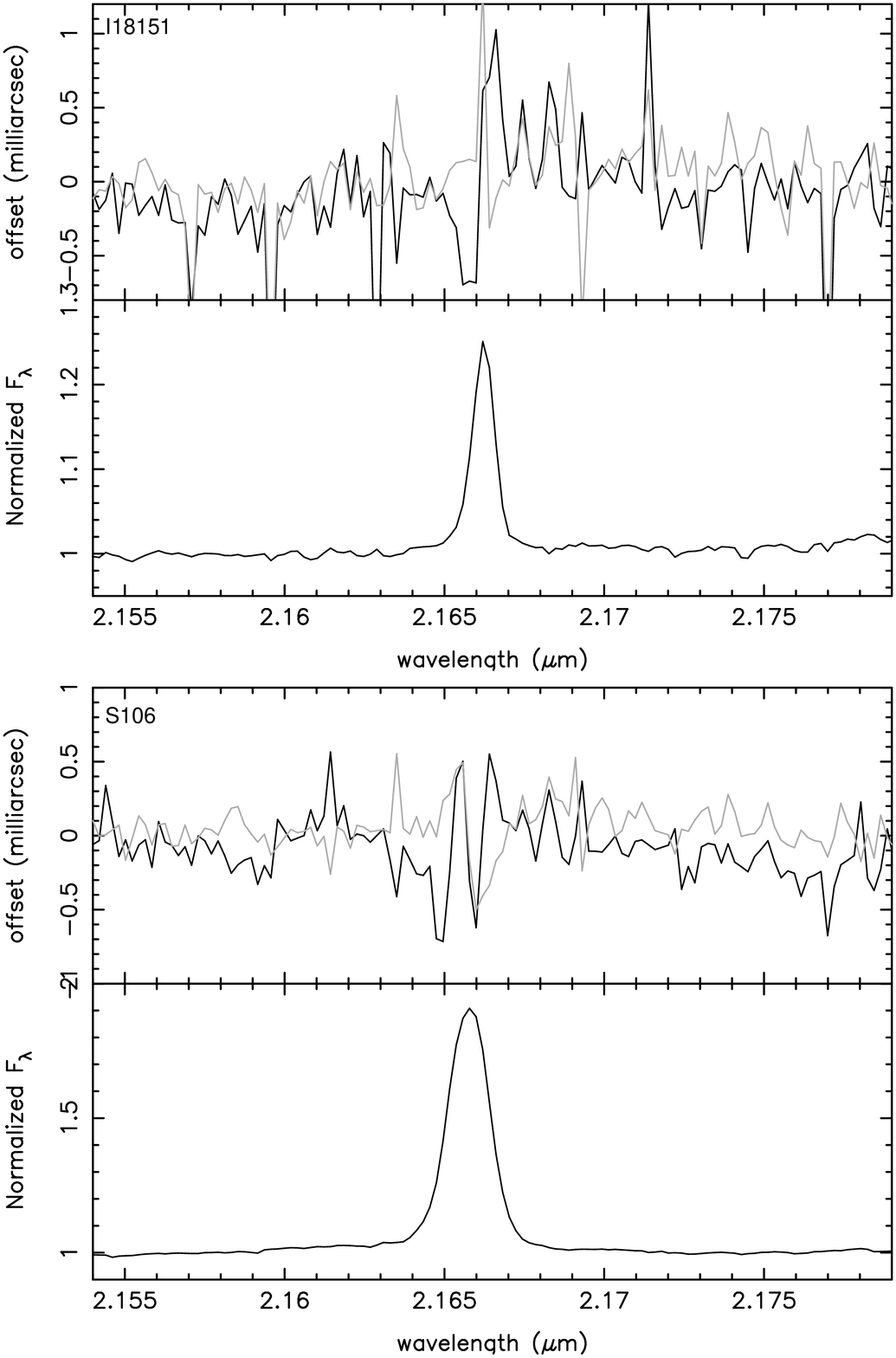}
  \caption{Spectro-astrometry of the Br$\gamma$ emission line for I18151
           (top panel) and S106 (bottom panel).  The black and grey lines
           are for the right ascension and declination, respectively.
         }
  \label{posacc}
\end{figure}

The spectral lines have a bell shape in general, with FWHM ranging from 
100 to 220~kms$^{-1}$.  These values are consistent with the outflow
velocities observed by \citet{bhd95}.  In V645~Cyg, we see a P-Cygni profile in
the velocity range between $-800$~kms$^{-1}$ and $-300$~kms$^{-1}$. The
Br$\gamma$ emission from AFGL~2136 is relatively weak compared to the other
sources.  The Br$\gamma$ luminosity has a positive correlation with the
accretion rate in lower mass stars \citep{muzerolle98,mendigutia}, hence the
weak Br$\gamma$ emission may be the result of a low accretion rate
\citep[e.g.][]{fe01,antonlucci11}.  However, this is not the only cause of weak
emission lines, as discussed more fully in \cite{cooper12}.  In particular,
there is no real evidence for underlying absorption in the Br$\gamma$ line as
is often seen in weakly accreting T Tauri stars.

Figure\,\ref{imgbg} presents the Br$\gamma$ emission line images.  They have a
similar appearance to the continuum images.  In AFGL~2136, the
\textsf{V}-shaped feature is not detected because of the low SNR.  The
nebulosities are somewhat extended with the typical FWHM sizes of 0.2~arcsec
apart from I19065 and V645~Cyg, in which the values are 0.13~arcsec and
0.1~arcsec, respectively.  These extensions are the same as those of the nearby
continuum.  These results suggest that the Br$\gamma$ emitting region is so
compact and close to the central star that they are not resolved with an 8~m
single aperture telescope and the nebulous features seen in the images are
light scattered by the circumstellar dusty envelope or disc.  This was also
concluded for the MYSO W33A (D10).

We performed three-dimensional spectro-astrometry to analyse the small-scale
geometry of the Br$\gamma$ emission.  Spectro-astrometry is a technique that
measures the positional centroid of the emission at an accuracy exceeding
the diffraction limit, see e.g. \cite{baines}.  D10 successfully revealed
a signature of a bipolar jet in the Br$\gamma$ emission line with a sub-mas
position offset between the blueshifted and redshifted components.
We consider velocity ranges between $-800$~kms$^{-1}$ and +400~kms$^{-1}$ for
V645~Cyg and between $\pm v_\mathrm{FWZI}$ for other objects.  In each velocity
(wavelength) channel, the astrometric centroid was calculated.  We also applied
the same procedure to the telluric standard stars to estimate the accuracy of
the position centroid and find 0.1 -- 0.4~mas.  Figure\,\ref{BG_spcast}
presents the results of the spectro-astrometry for all of our sources across
the Br$\gamma$ emission line.  The positional offsets are shown with respect
to the location of the continuum emission.

%

In most cases, the trails seem to show random walk like patterns.  The random
walk pattern as seen in I19065 covers a larger range than in the other such
sources, and is especially pronounced near the systemic velocity.
This is consistent with the classification of this object in the RMS database
as a weak, and very compact, HII region.  AFGL~2136 shows a strong excursion
near the systemic velocty, which may also be evidence for a weak extended and,
in this case, asymmetric HII region.

The strongest coherent features however are shown by I18151 and S106.
Figure\,\ref{posacc} shows the centroid positions of right ascension and
declination of these objects, which are indicated with black and red lines,
respectively. In both cases, the position centroids vary randomly at the
nearby continuum, with an error of about 0.1mas, consistent with the accuracy
as measured on the telluric standard stars. However significant excursions
can be seen across the line Br$\gamma$ line itself, consistent with an
east-west signature in I18151 for example, as also shown in
Fig.\,\ref{BG_spcast}.  Hence we conclude that the spectro-astrometry traces
real effects across the emission line.

In S106, the highest velocity points are separated by $\sim0.4$~mas with a
position angle of $\sim100\degr$ east of north.  This agrees reasonably well
with the observed orientation of the radio emission at the core ($\sim 119$
east of north) as shown by \citep{hoare94}.  Both of these lie perpendicular
to the orientation of the ionised bipolar lobes seen in images of the more
general HII region around S106 \citep{hoare94}.  Therefore our
spectroastrometric signature is consistent with a source in the ionised gas
in, or arising from, a circumstellar disc (including a possible disc-wind).
The inner part of the bipolar lobes are themselves traced in the
spectroastrometric data near systemic velocities by the large positional
offsets across a range of angles centred on a NNE--SSW direction.

The offset of $\sim$0.4~mas between the blue and red components corresponds to
0.68~AU at 1.7~kpc.  The velocity of the gas at these points is approximately
$\pm200$~kms$^{-1}$ around the systemic velocity.  Assuming Keplerian rotation,
and given the almost edge-on orientation of 83$\degr$ (see Section
\ref{cobh}), we estimate a central mass of $\sim22\pm5M_{\sun}$ (where the error
is solely that from the spectroastrometric accuracy).  This is
consistent with the inferred spectral type and luminosity as given in Table
\ref{obs_parm}.  It should be noted that the true separation between the blue
and red components of the rotating material is likely to be larger than the
aforementioned value. The contribution to the total flux from the line emission
of the rotating material is smaller than that of the continuum and the stellar
component for all our sources.  The centroid of the total spectrum therefore
lies nearer to the continuum than a ``pure line'' spectrum would be.

In I18151 a similar pattern is seen at high velocity, with a position angle of
approximately 90$\degr$.  A larger scale collimated jet detected in molecular
hydrogen extends from southeast to northwest (D04).  We see the southeast part
of this jet clearly, as shown by Fig.\,\ref{h2_i18151}.  As noted previously
however, it appears that the H$_2$ jet initially starts in a much more
southerly direction within the inner 0.5 arcseconds, perpendicular to the
spectroastrometric signature seen in Br$\gamma$.  We therefore again ascribe
the source of the Br$\gamma$ emission to a circumstellar disc, or disc-wind. 

The size of the disc is less clear from the data for I18151 than it is for
S106.  However, the same analysis as for S106, assuming an inclination angle of
66$\degr$ (again see Section \ref{cobh}), gives a lower limit to the mass of
$\sim10\pm5M_{\sun}$.  This is again consistent with the known properties of
this source.

\begin{table*}
  \begin{center}
  \caption[]{The fitting results of the CO bandhead emission feature. The
      second line for each object indicates the range of parameter space
      searched over.  The errors were calculated by holding all but one
      parameters constant at their best fitting value, and varying the selected
      parameter until the difference in $\chi^2$ was equal to unity.  The error
      of $R_\mathrm{out}$ is not presented because $R_\mathrm{out}$ is defined
      as the radius at which the temperature drops to 1000~K.  Below this
      temperature, the $\nu=2-0$ emission is no longer excited.  The error
      values indicated with an asterisk are not well determined because they
      lie at the boundary limit of the scanned ranges.}
  \label{tab:cobh}
  \begin{tabular}{llllllll}
  \hline
  \hline
  \multicolumn{1}{c}{target} & \multicolumn{1}{c}{$R_\mathrm{in}$} &
  \multicolumn{1}{c}{$R_\mathrm{out}$} & \multicolumn{1}{c}{$i$} &
  \multicolumn{1}{c}{$T_0$} & \multicolumn{1}{c}{$\Delta v$} &
  \multicolumn{1}{c}{$\log N_\mathrm{CO}$} &
  \multicolumn{1}{c}{$\chi^2$} \\
         &  \multicolumn{1}{c}{AU} &  \multicolumn{1}{c}{AU}  &
  \multicolumn{1}{c}{deg} &  \multicolumn{1}{c}{K}  &
  \multicolumn{1}{c}{kms$^{-1}$} &
  \multicolumn{1}{c}{cm$^{-2}$}  &      \\
  \hline
  I18151 & 2.4$^{+5.8}_{-1.4}$ & 28.6 &  79$^{+11.3*}_{-49*}$ & 4400$^{+600*}_{-2250}$ & 5.3$^{+4.2}_{-5.2*}$ & 21.4$^{+5.4}_{-17}$ & 8.35 \\
  \hline
         & 0.37 -- 8.0        &       & 30.0 -- 90.0 & 1000 -- 5000 & 0.1 -- 50.0 & 1.0 -- 50.0 &  \\
  \hline
  S106   & 0.3$^{+0.26}_{-0.11}$ &  4.0 &  89.7$^{+0.3*}_{-39*}$ & 4600$^{+400*}_{-1800}$ & 1.2$^{+5.3}_{-1.1*}$ & 22.7$^{+17*}_{-1.8}$ & 1.74 \\
  \hline
         & 0.1 -- 5.8           &       & 50.0 -- 90.0 & 1000 -- 5000 & 0.1 -- 50.0 & 4.0 -- 50.0 &  \\
  \hline
  V645~Cyg   & 3.2$^{+0.7}_{-2.8*}$ & 44  &  10.0$^{+25}_{-10*}$ & 4900$^{+100*}_{-900}$ & 46.6$^{+29}_{-34.8}$ & 20.0$^{+0.7}_{-15}$ & 3.48 \\
  \hline
         & 0.1 -- 5.0           &      & 0.0 -- 50.0 & 1000 -- 5000 & 0.1 -- 100.0 & 1.0 -- 50.0 &  \\
  \hline
  \end{tabular}
  \end{center}
\end{table*}

\subsection{CO $\nu=2-0$ lines}\label{co}
\subsubsection{Bandhead emission}\label{cobh}

\begin{figure}
  \centering
  \includegraphics[width=8.5cm]{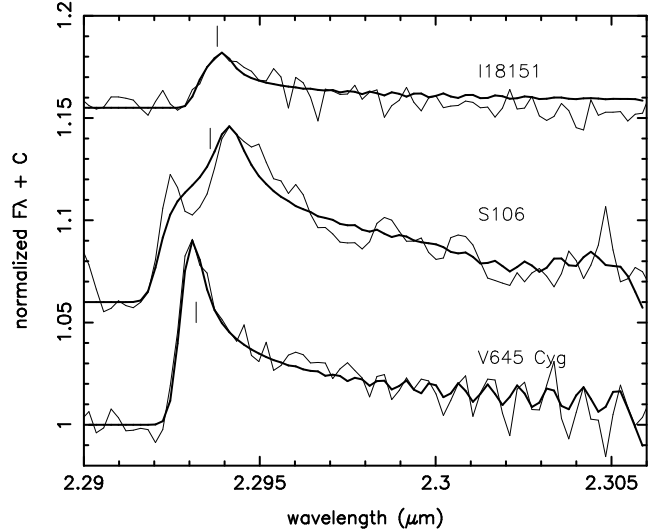}
  \caption{CO $\nu$=2--0 bandhead spectra. The thin and thick lines denote
           the observations and the model results, respectively.  The bars
           around $\lambda=2.293~\mu$m indicate the local standard of rest.
         }
  \label{CO_bh}
\end{figure}

In I18151, S106, and V645~Cyg, the CO $J=2-0$ bandhead emission is detected
(as previously seen in S106 by \citet{chandler95} and in V645~Cyg, as discussed
in \citet{clarke06}).  Fig.\,\ref{CO_bh} presents the spectra.  S106 shows
a clear double-peaked profile at the bandhead.

The interpretations of the spectral shape of the bandhead have been provided by
several authors \citep{chandler95,najita96,kraus00}.  Figure\,2 in the paper by
\cite{najita96} is instructive to explain the spectral features of a disc
model.  The emission of a single line from an inclined Keplerian rotating disc
is described with the double-horned rotational broadening function.  The model
bandhead is then simply the convolution of this broadening function with the
intrinsic CO emission.  This leads to a distinct ``blue shoulder'' of the
bandhead where there is blue shifted emission relative to the line of sight,
which sets a constraint on both the line width and the inclination of the
source.
This is clearly seen in S106.  It should be noted that a double-peaked feature
can also be reproduced with a wind model \citep{chandler95}, when the opening
angle of the conical wind is so small that the observer sees the red component
only from the far side and the blue component only from the front side.
This is not the case in S106 because this object is seen near to edge-on
\citep{bs82,schneider07,saito09}.

The spectral features of V645~Cyg can also be explained with the same models.
The emission peak is located around 2.293~$\mu$m, which is between the blue
shoulder and red peak in S106 and where the difference of the system velocities
is negligible.  In the face-on case, the rotational broadening function is
single-peaked, and the peak stays at the bandhead wavelength.  In this object,
only the blueshifted components are detected in the \mbox{[S\,{\sc ii}]} and
\mbox{[O\,{\sc i}]} outflow emissions \citep{hp89,acke05}.  This result
indicates that the system is seen near face-on and the redshifted part is
obscured by the equatorial material, presumably an accretion disc.
A single-peak is also reproduced in a wind model if the opening angle of
the wind is large.  However we would then expect the peak of the emission
to be shifted redwards compared to the nominal $v_{LSR}$ of the source and
it clearly is not. 

For I18151, the data, although noisy, show no significant emission bluewards
of the peak of the bandhead that cannot be explained by simple line broadening.
The inclination angle of I18151 is estimated to be $60\degr$
\citep{fallscheer11}, intermediate between the cases of S106 and V645~Cyg, and
the appearance of the spectrum is consistent with this if the origin of
the emission is a disc.

\begin{figure*}
  \centering
  \includegraphics[width=12cm]{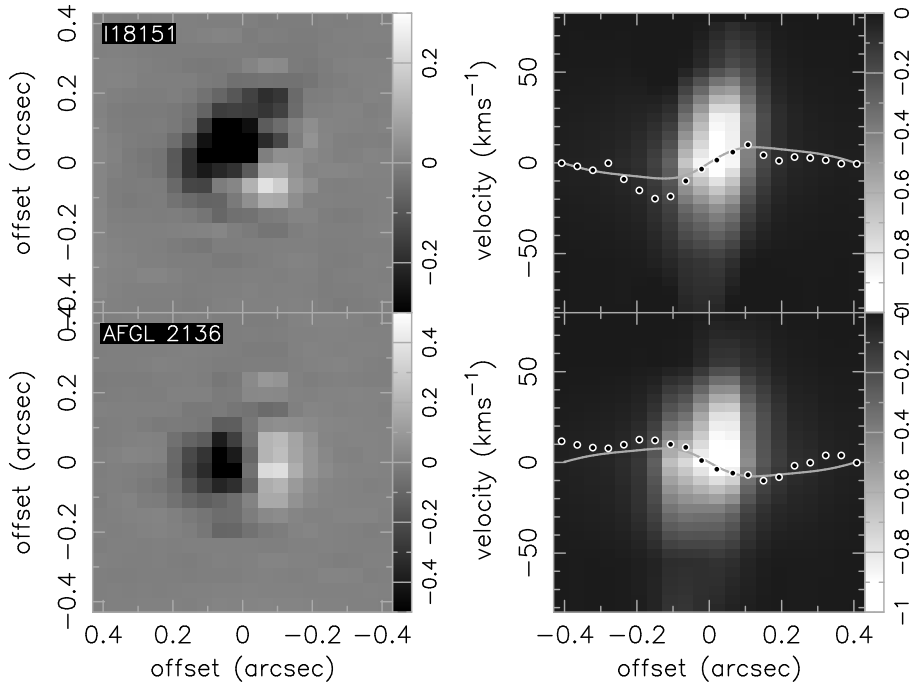}
  \caption{{\it Left:} the differential image between the blueshifted and
    redshifted components of the CO absorption features, i.e.\,the subtraction
    of the blue component flux image from the red one.  {\it Right:} the
    position-velocity map at the inner 20$\times$20 pixel
      (0.86$\times$0.86~arcsec) region.  The top and bottom panels are the
    results of I18151 and AFGL~2136, respectively.  The white circles are
    the velocity centroids of the observed data.  The grey lines denote
    the Keplerian velocity plot convolved with Gaussian functions with the PSF
    sizes, where the central masses of 30~$M_{\sun}$ and 20~$M_{\sun}$ are
    assumed for I18151 and AFGL~2136, respectively.
    }
  \label{CO_vc}
\end{figure*}

We have also used a simple model of the gas phase CO disc to fit the data.
We analyze the spectra attributed to the CO first over-tone bandhead emission
based on \cite{kraus00} to derive the physical parameters of the disc geometry
\citep[see also][]{wheelwright10,ilee12}.  In our model, up to $J=100$
rotational levels for the $\nu=2-0$ under the local thermodynamic equillibrium
are considered.  The CO energy levels are calculated with the $Y_{k,l}$
parameters \citep[see equation (6);][]{kraus00,farrenq91}, and the Einstein
coefficients are taken from \cite{chandra96}.  The CO$/$H$_2$ ratio is assumed
to be $10^{-4}$.  For the disc geometry, we use a temperature gradient model,
$T\left(r\right)=T_0\left(r/R_\mathrm{in}\right)^{-p}$, and a geometrically
thin disc with a surface density
$\Sigma\left(r\right)=\Sigma_0\left(r/R_\mathrm{in}\right)^{-q}$, where $r$ is
the distance from the central star.  The disc has inner and outer radii of
$R_\mathrm{in}$ and $R_\mathrm{out}$, respectively, and is tilted with an
inclination angle, $i$, where $i=0\degr$ indicates face-on.  The CO molecules
are assumed to have a Keplerian rotating motion
$v_\mathrm{CO}=\sqrt{GM_{\star}/r}$, where $G$ and $M_{\star}$ are the
gravitational constant and the stellar mass, respectively.  The disc is
divided into 75 radial rings each with 75 azimuthal cells.  The intrinsic line
shape of each transition has a Gaussian function with a line width of
$\Delta v$.  The total spectra are obtained by summing the flux from
individual cells, in which the wavelength shift due to the rotational velocity
of the disc is taken into account.  The outer radius is set by the condition
that the gas phase temperature drops below 1000~K, at which point the $\nu=2-0$
emission is no longer excited.  In addition, the C-O bonding is dissociated at
temperature above 5000~K.  Hence, the maximum temperature is chosen not to
exceed 5000~K.  With these physical assumptions the CO bandhead spectra are
produced and compared to the observed spectra using an IDL script.
The best fit model parameters are found using $\chi^2$ minimization.
We examine the three objects with CO emission, I18151, S106 and V645~Cyg.
The distance and the stellar mass are assumed to be 2.0~kpc and 15~$M_{\sun}$
for I18151, 1.7~kpc and 20~$M_{\sun}$ for S106, and 3.6~kpc and 20~$M_{\sun}$
for V645~Cyg (see Table \ref{obs_parm}).  With these assumptions, we searched
for the best fit parameters of $R_\mathrm{in}$, $R_\mathrm{out}$, $p$, $q$,
$i$, $T_0$, $\Delta v$, and the column density of CO,
$N_\mathrm{CO}\left(=\Sigma_\mathrm{0,CO}\right)$.  It turned out that the $p$
and $q$ parameters have particularly large uncertainties.
Hence, we set fixed values of $\left(-0.6, -1.5\right)$, respectively,
which are derived for intermediate mass stars \citep{ilee12}.
The results of other six parameters are presented in Table\,\ref{tab:cobh}.  
Only $i$ and $T_0$ are estimated with good accuracy.  A similar analysis was
done for S106 by \cite{chandler95}.  Although the estimated inclination angle
they used of $i=65\degr$ is somewhat different from ours, their estimate of 
the size of the CO emitting region of 0.8 to 1.5~AU is in reasonable agreement.
In both S106 and V645~Cyg, the best fitting inclination angle is consistent
with the qualitative expectations discussed previously, as well as available
results from the literature as given in Appendix \ref{discussion}, suggesting
that our estimated stellar masses are also reasonable.

\subsubsection{The low $J$ absorption transitions}\label{CO_lowJ}

D10 analysed the low J absorption line velocity structure of CO $\nu=2-0$.
Their data show a pair of blueshifted and redshifted components at east and
west with respect to the flux peak position in the $K$-band continuum in W33A.
They found the classical signature of a Keplerian rotating structure in the
position-velocity (PV) diagram and a central mass of 15~$M_{\sun}$, which
includes the masses of the central star and the hot gas inside the cool gas
region.

Both I18151 and AFGL~2136  show absorption features in the low $J$ transition
of CO $\nu=2-0$.  From the cube data, we used 11 absorption lines, which appear
between 2.32~$\mu$m and 2.35~$\mu$m \citep{bw84}.  We first derived the
wavelength positions of individual absorption features, which are regarded as
the zero-velocity.  The images of all branches are aligned with the relative
velocity and are added to increase the signal-to-noise ratio.  Then,
the differential images between the redshifted
($v>0$) and blueshifted ($v<0$) components, i.e.\,the subtraction of the
blueshifted component image from the redshifted component images, the images
of the velocity centroids, and the PV diagrams are made.

The top and bottom panels of Fig.\,\ref{CO_vc} show the results of I18151 and
AFGL~2136, respectively.  The velocities range between
$\pm15\left(\pm10\right)$~kms$^{-1}$ for I18151 and
$\pm10\left(\pm7\right)$~kms$^{-1}$ for AFGL~2136.  Although the standard
deviations in velocity are large, we can clearly see a pair of redshifted and
blueshifted components in the $B-R$ images (left panels).  The separations and
the position angles of these components are obtained by measuring the
absorption centroids of the blueshifted and redshifted components to be
0.038~arcsec at $+31\degr$ for I18151 and 0.054~arcsec at $+91\degr$ for
AFGL~2136 with typical errors  of one tenth of a pixel on the centroid  and
5$\degr$ on the angle.  Applying a slit along these position angles, the PV
diagrams are made (right panels) and the velocity centroid at each position
pixel (0.043~arcsec spacing) is plotted with pink dots, with a typical error
on the velocity of better than one tenth of a resolution element,
or kms$^{-1}$.  The redshifted component is located on the right side of
the panel.  In the velocity centroid maps, we see a skew, which is often seen
in objects with Keplerian rotation, though the correspondence with a Keplerian
model is not perfect even given the estimated errors.  If we assume that this
is also the case in our results, we can estimate the central masses.
For this purpose, we made a simple model.  The Keplerian rotating velocity of
the CO gas, $v_\mathrm{CO}$, is calculated by $v_\mathrm{CO}=\sqrt{GM/r}$,
where $M$ is the central mass.  The inclination angles, $i$, and the distances
to these objects are assumed to be $60\degr$ and 2.0~kpc for I18151
\citep{fallscheer11} and $70\degr$ and 2.0~kpc for AFGL~2136
\citep{kastner92,murakawa08,dewit11}.  To allow direct comparison with the
observations, the resulting velocity curve is obtained by convolving Gaussian
functions with the PSF sizes (0.1~arcsec).  In the PV diagram, the modeled
velocity curves are plotted with green lines.  The observed and model data are
sufficiently similar to conclude that we do see Keplerian rotation in the cool
absorbing gas in these sources.  The resultant central masses are 30~$M_{\sun}$
for I18151 and 20~$M_{\sun}$ for AFGL~2136, with error of approximately 10\%
assuming the Keplerian model is a good fit.

The central mass of 30~$M_{\sun}$ for I18151 is larger than that of the
inferred mass for a B0 star of $15M_{\sun}$ that was used in modelling the CO
bandhead emission.  However, the absorbing gas includes the mass of not only
the central star but also all the gas located inside the maximum extent of the
region of absorbing gas.  Although the error on these measurements is large
enough to suggest both absorption and emission kinematical masses could be
equivalent, it may also suggest a larger reservoir of material still available
for accretion by the central star.  The large 220$M_{\sun}$ mass found for the
``disc-like'' structure by \cite{fallscheer11} is certainly consistent with the
picture in which more accretion could take place.  A similar situation was
found by D10 for W33A as well.

The orientation of the redshifted and blueshifted components in the CO
$\nu=2-0$ absorption line is nearly northeast-southwest in I18151,
perpendicular to the larger scale outflow.  This is notably different from both
the ionised spectroastrometric signature seen for Br$\gamma$, and the small
scale north-south H$_2$ jet we see in our data.  It is possible that the
outflow is precessing with time, though no intermediate angles are seen in the
H$_2$ jet by D04.  It is also possible that the jet initially launches in a
north-south direction, but is deflected by the larger scale molecular envelope
in the observed southeast-northwest direction.  In either case, the structure
of the accreting gas in the immediate vicinity of the star appears to lie at
different orientations when moving further away from the star.

\begin{figure}
  \centering
  \includegraphics[width=8.5cm]{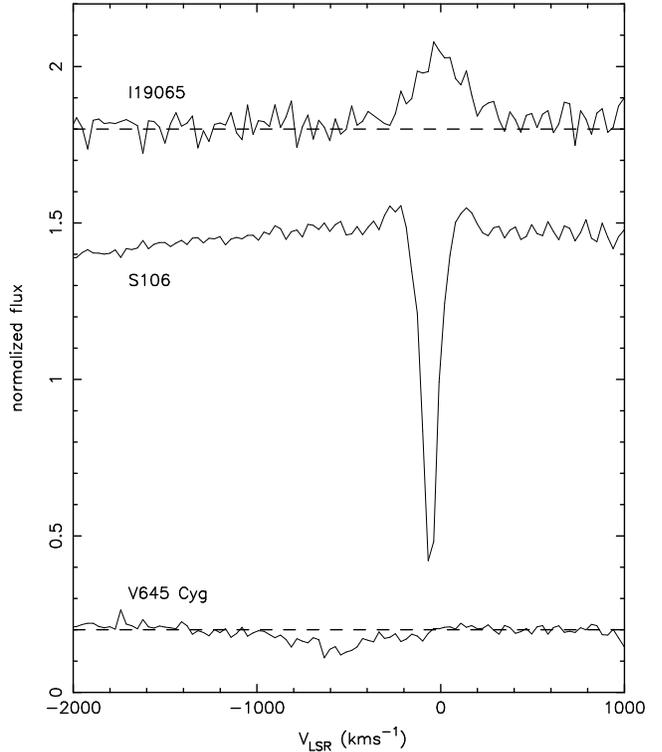}
  \caption{Spectral features of the \mbox{He\,{\sc i}} at 2.05689~$\mu$m
           on velocity scale.
         }
  \label{heI}
\end{figure}

The orientation of the CO absorption in AFGL~2136 is also puzzling.  The blue-
and red-shifted components of the absorption centroid lie at a position angle
of $90\degr$ with a separation of $\sim$100~AU.  As noted in Appendix
\ref{discussion}, the observed outflow axis runs SE--NW, consistent with the
polarisation disc with a position angle of $\sim$45$\degr$ found by
\cite{murakawa08}.  It is possible that the distribution of mass on a few
thousand AU is different from that on hundreds~AU, i.e.\, a simple axisymmetric
structure may not be applicable in this object, just as appears to be the case
for I18151.  Nevertheless, by assuming a Keplerian rotating motion in the cold
CO gas, the velocity structure can be reproduced well with a 20~$M_{\sun}$ mass
model.  Although the stellar parameters of AFGL~2136 have not been precisely
determined, it corresponds to a late O zero-age main sequence star in terms of
the estimated luminosity of 5.3$\times$10$^4~L_{\sun}$, corresponding to a
$20-25~M_{\sun}$ star.

\subsection{\mbox{He\,{\sc I}} and other ionised and atomic emission lines}\label{helium}
In our sample, the \mbox{He\,{\sc i}} $2^1P-2^1S$ ($\lambda=2.05869~\mu$m) line
is detected in I19065, S106, and V645~Cyg, and the $4^{1,3}S-3^{1,3}P$
($\lambda=2.1128/2.1137~\mu$m) blend is present as a weak absorption feature in
S106.  Figure\,\ref{heI} presents the line profile on a $v_{LSR}$ scale for the
$2^1P-2^1S$ in all three objects.  We see emission in I19065 and absorption in
S106 and V645~Cyg.  Notably the emission feature seen in I19065 is broader than
expected from unresolved emission from the ionised HII region, suggesting that
this, like the Br$\gamma$ emission, arises from the star in this source.

\cite{cs00} have obtained $K$-band spectra of Be stars, and \cite{hanson96} and
\cite{hanson05} of a more general sample of OB stars of all classes.  The
trends with spectral type are relatively clear and can be used to classify the
emission as seen in our targets.  HeI $2^1P-2^1S$ appears in emission in later
type stars in dwarfs, giants and supergiants, and in many of the B0Ve and B1Ve
stars in \cite{cs00}, and in Oe stars with types as early as O6Ve in
\cite{hanson96}.  The lack of any detectable $4^{1,3}S-3^{1,3}P$ line in I19065
however suggests a relatively narrow range of type of around B0Ve or B1Ve,
which is consistent with the known luminosity of this source from the RMS
survey \citep{mottram}.  

\citet{dbh93} previously detected the $2^1P-2^1S$ line in S106, mostly in
absorption as we also see it.  They argue that this absorption probably arises
due to high optical depth in the 584~\mbox{\AA} line and a high electron
density ($n_e\ga10^8$~cm$^3$).  The FWHM is 65~kms$^{-1}$ and is in good
agreement with the result of 70~kms$^{-1}$ by \cite{dbh93}.  This velocity
width is significantly smaller than that of the Br$\gamma$ emission lines
($\sim$100 -- 200~kms$^{-1}$).  \cite{dbh93} proposed that this line traces the
region where the outflow from a disc-wind is still accelerating.  Weak HeI
$4^{1,3}S-3^{1,3}P$ absorption is seen, but there is no NIII 8--7 emission at
$\sim2.116~\mu$m emission as would be seen in earlier spectral types
\citep{hanson96}.  This suggests a type of O9Ve--B0Ve. We also carried out
a spectroastrometric study of the $2^1P-2^1S$ line given its strength.
Overall there is only a weak signature near the systemic velocity, in the
direction of the bipolar nebula, perhaps consistent with a slight degree of
infilling of the absorption line profile by larger scale nebular emission.
Our lower resolution spectroscopy taken for the RMS survey \citep{cooper12}
shows nebular HeI $2^1P-2^1S$ emission consistent with this.
The spectroastrometry does not show evidence for any other feature.
In particular, there is nothing similar to the disc like structure seen in
Fig.\,\ref{BG_spcast}.

In V645~Cyg, the absorption feature is strongly blueshifted and appears around
$v\sim$$-800$~kms$^{-1}$.  A similar trend is seen in the spectrum presented by
\cite{clarke06}.  However, the spectral feature spreads up to
$-2000$~kms$^{-1}$ in their data, but is detected only up to $-1000$~kms$^{-1}$
in our spectrum.  V645~Cyg changes its spectral appearance regularly,
complicating any attempts to derive an exact spectral type.  However, the
spectra we have is consistent with approximately O9Ve, similar to S106.  The
star cannot be hotter, given the lack of NIII, and the lack of any HeI
$4^{1,3}S-3^{1,3}P$ suggests it cannot be much cooler either.

The other metal lines that we see in our spectra, such as NaI, MgII or FeII are
less sensitive to the spectral type.  In all three cases it is plausible
 that the
emission actually comes from the accretion disc rather than the stellar
atmosphere \citep{lumsden12}.  

Overall the HeI and other metal lines suggest a common picture in which a
stellar wind or disc wind can be present, along with an accretion disc.  It
would be interesting to determine whether spectroastrometry could be used on
some of these metal lines in future higher signal-to-noise studies to test
whether this scenario is correct.


\section{Conclusions}
We presented high-resolution $K$-band integrated field spectroscopy of six
MYSOs obtained using the AO assisted NIFS instrument on the Gemini-north
telescope.  This observing technique allows us to analyse three-dimensional
spectroscopy across emission and absorption lines.  In the previous paper
(D10), we successfully detected a bipolar jet like feature in the Br$\gamma$
emission line and a Keplerian rotating motion of the neutral CO gas disc around
a $\sim$15~$M_{\sun}$ mass central star in W33A.  In this new work, we increase
the sample, with four out of six being already well studied and two new.  Our
new data display a number of spectral lines such as Br$\gamma$ emission lines,
CO bandhead emissions and absorptions, H$_2$ $\nu=1-0$ S(1), S(2), and
\mbox{He\,{\sc i}}.  Some clear features can be inferred from our data despite
the heterogeneous nature of the sample.

All of our targets show evidence of Br$\gamma$ emission.  The measured
$v_\mathrm{FWHM}$ ranges between 100 and 200~kms$^{-1}$, which is consistent
with stellar wind or disc-wind velocities \citep{bhd95}.  In V645~Cyg, the
P-Cygni profile is detected in a velocity range between $-800$~kms$^{-1}$ and
$-300$~kms$^{-1}$.  This is consistent with a spherical wind.  In the
Br$\gamma$ emission images, the appearance is similar to those of the $K$-band
continuum images in all objects, suggesting that these features are seen in
scattering by dust in the circumstellar disc or envelope.  

There is a clear spectroastrometric signature in both S106 and I18151, running
perpendicular to the known larger scale outflow axis in both cases.  At least
for S106, the orientation is consistent with the disc-wind picture previously
developed for this source, and we conclude the same should be true for I18151
as well.  This is the opposite of what D10 saw in W33A, where the best
explanation for the spectroastrometric signature was an ionised bipolar jet.
In the two cases presented here it would suggest that the signature arises in
ionised gas in, or from, a disc around the star.  In the case of the well
studied S106, the orientation of our spectroastrometric disc is also consistent
with the morphology of the radio emission observed by \cite{hoare94}.  The
presence of a stellar wind in this object is clear from the strong HeI
absorption, lending support to the disc-wind picture of \cite{dbh93}.  The
estimates that are derived for the stellar masses from the spectroastrometry
assuming Keplerian orbits, although clearly imprecise due to the scatter in the
data as shown in Fig.\,\ref{BG_spcast}, and strictly lower limits, are
consistent with the known properties of these sources.  The evidence that these
two sources contain discs is therefore strong.

The other objects show spectroastrometric data consistent with a null result in
most cases, revealing only a random-walk like pattern, though AFGL~2136 may
show evidence for a weak HII region near the systemic velocity.  I19065 shows a
larger scatter in its random-walk pattern than the other non-detections.  This
may be due to ``contamination'' of the stellar spectrum due to the extended
nebular HII region emission.  The lack of any signature in V645~Cyg, given the
excellent signal-to-noise for that source, places a strong constraint on what
we see there, consistent with the P-Cygni wind profile observed in the line.

Evidence for a disc from CO bandhead emission was present in I18151, S106 and
V645~Cyg.    The disc parameters for I18151 and S106 are consistent with those
found from our analysis of the spectroastrometry of Br$\gamma$.   The edge-on 
nature of S106 is clearly seen in these data as well.  V645~Cyg shows evidence
for a more pole-on orientation, which may be why we see no evidence for a 
disc-wind component in this source in the spectroastrometry.   It may also
simply be that V645~Cyg is relatively more evolved than the other sources.

Finally, in I18151 and AFGL~2136, a series of CO absorption lines are detected.
If we assume that these lines originate from the CO gas disc and the velocity
structures reflect the Keplerian rotating motion, the central masses can be
estimated.  Our data show a separation of the flux centroids between the
blueshifted and redshifted components.  This orientation is nearly
perpendicular to the jet direction in I18151, giving a strong evidence for
a rotating motion in our data, whereas that orientation is less correlated
with the direction of the outflow or the NIR polarization disc in AFGL~2136.
Nevertheless, the rotation velocities and the central masses are estimated
to be $\pm15(\pm10)$~kms$^{-1}$ and 30~$M_{\sun}$ for I18151 and
$\pm10(\pm7)$~kms$^{-1}$ 20~$M_{\sun}$ for AFGL~2136.

Our work has demonstrated that three-dimensional spectroastrometry across
spectral emission and absorption lines is a powerful diagnostic technique
to study the geometrical and kinematic properties of ionized or molecular
regions in vicinity of the central star.  The ionized hydrogen traces the mass
accretion channel inside the accretion disc and disc winds, which is driven
by the central star.  On the other hand, the molecular gas probes the Keplerian
rotating discs.  The consistency of the results obtained using different
methods strongly supports the hypothesis that massive stars form by
mass accretion via circumstellar discs.

\section*{Acknowledgments}
Based on observations (proposal ID: GN-2011A-Q-56) obtained at the Gemini
Observatory, which is operated by the Association of Universities for Research
in Astronomy, Inc., under a cooperative agreement with the NSF on behalf of
the Gemini partnership: the National Science Foundation (United States), the
Science and Technology Facilities Council (United Kingdom), the National
Research Council (Canada), CONICYT (Chile), the Australian Research Council
(Australia), Minist\'{e}rio da Ci\^{e}ncia, Tecnologia e Inova\c{c}\~{a}o
(Brazil) and Ministerio de Ciencia, Tecnolog\'{i}a e Innovaci\'{o}n Productiva
(Argentina).  We thank the referee, Dr. Scott Wolk, for his helpful
comments.

\appendix
\section{Properties of individual objects}\label{discussion}
\subsection{I18151}
I18151 ($D=3$~kpc) has been studied well, having featured in a large number of
survey observations of MYSO candidates \citep{sridharan02,beuther02}.  Images
at the MSX 8~$\mu$m \citep{marseille08} and 1.2~mm continuum \citep{beuther02}
wavelengths revealed four clumps MM1--4 spreading across $\sim$$1\arcmin$ scale.
IRS~1 is located at MM1, which we observed.  \cite{fallscheer11} performed
a radiative transfer modeling of the dust disc of this object and obtained a
luminosity of $\sim16000~L_{\sun}$.  However, the spectral energy distribution
was not well constrained in the far infrared in their model.  Unpublished
Herschel PACS data at 70$\mu$m reveals a source with flux $\sim765$Jy.
\cite{fallscheer11} scaled all ``large beam'' flux down by a factor of four
to match their mm interferometry.  If we adopt the same correction, their best
fit spectral energy distribution lies a factor of $\sim1.7$ too low.
The Herschel data is only about 10\% smaller than the published large beam
IRAS flux we used to derive a luminosity of $22000~L_{\sun}$ for the RMS
database.  It seems likely the luminosity is actually $\sim20000~L_{\sun}$
given these remarks, correspondingq to a $15M_{\sun}$ star \citep{fallscheer11}.
\cite{fallscheer11} also found a huge disc of 220~$M_{\sun}$ mass and 5000~AU
radius from the mm interferometry.  Only the inner 30~AU satisfy the Toomre
criterion, a condition of gravitational stability \citep{toomre64}.  Hence,
most of the mass of their modeled disc would include the outer envelope.

Our results are consistent with the luminosity and mass for the central star
implied by these earlier results.  The disc orientation we find in Br$\gamma$
is east-west, whereas the large scale H$_2$ jet traced by D04 lies NW--SE.  The
CO absorption agrees with the latter orientation.  This suggests that the inner
disc lies at an angle to the larger scale envelope structure.

\subsection{AFGL~2136}
Although a couple of stars or cores are often found in a high-mass star
formation region, there is no other source within $5\arcmin$ from AFGL~2136 in
the mid-infrared and submillimeter wavelength ranges \citep{urquhart09}.  The
near-infrared images show three features with extensions of a few arcsec and
the central proto-star is located at IRS~1 \citep{kastner92}.  The CO emission
line observations show a bipolar outflow extending at a position angle of
$\sim$135$\degr$ and the SE and NW components are blueshifted (14 to
20~kms$^{-1}$) and redshifted (24 to 30~kms$^{-1}$) respectively.  The
orientation of the polarization disc in the NIR data is found to be
$\sim$45$\degr$, approximately perpendicular to the CO outflow direction
\citep{minchin91,murakawa08}.  These results indicate that the orientation of
the equatorial plane lies at a position angle of $\sim$45$\degr$.  The results
of radiative transfer modeling suggest that the viewing angle is about
$70\degr$, i.e.\,seen near edge-on \citep{murakawa08,dewit11}.  \cite{dewit11}
found that a situation requiring both a gaseous disc and dusty envelope is
needed to explain the observed mid-infrared interferometry.  The observed
baseline for their data lay along the position angle of the polarization disk.

\cite{mv04} detected both water and class II methanol maser spots about
1~arcsec west with respect to IRS~1.  These masers are usually tracers of
outflows, but they do not align with the CO outflow presented by
\cite{kastner92}.  Instead they lie perpendicular to the plane of the CO
absorption feature we find.  The strongest water maser feature found by
\cite{mv04} is redshifted with respect to the systemic velocity.  Our more
recent water maser observations using the GBT \citep{urquhart11} shows that
this feature has weakened considerably, with the strongest component now
arising much closer to the systemic velocity.  

\cite{mv04} concluded that this may be an example where the water maser arises
in the infalling gas, in the post-shock gas just behind the accretion shock.
It may be that the CO absorption disc is evidence for larger accretion flow
that is ``twisted'' relative to the small scale gaseous disk suggest by
\cite{dewit11}, giving rise to a stronger shock than would otherwise be the
case, and reconciling the position angles of the small scale disc and larger
scale absorbing gas.  Further observations are clearly required to fully
resolve this puzzle.

\subsection{S106}
The inner circumstellar region of S106 has been studied by several authors
\citep[e.g.][]{bs82,hoare94,chandler95,ghosh03,schneider07,saito09}.
The dark lane is thought to be an edge-on massive disc \citep{bs82}.
\cite{lucas78} found a rotation axis of $\sim$$30\degr$, which is perpendicular
to the dark lane.  The radio spectrum at 22~GHz shows a flux, $S_\nu$,
proportional to $\nu^{0.7}$, suggesting an ionized stellar wind \citep{pf75}.
The wind has a high-mass loss rate of $10^{-5}~M_{\sun}$yr$^{-1}$
\citep{hm81,felli84}.  This is too high for OB type main sequence stars,
indicating a young stellar object.  The disc plane is well defined by the radio
observations of \cite{hoare94} to be $\sim120\degr$.  

Line observations have shown strong P-Cygni profiles in hydrogen recombination
and \mbox{He\,{\sc i}} \citep{dbh93} and double peaks in Br12 and
\mbox{Fe\,{\sc ii}} \citep{lumsden12}.  These features can not be explained
with an accretion flow, but suggest emission from stellar$/$disc winds or an
accelerating outflow.  This can be see as consistent with the fact that the
radio emission lies in the disc plane and not perpendicular to it if the
source of both is a disc-wind type geometry \citep{hoare94}.  Our own
Br$\gamma$ spectroastrometry is consistent with these previous findings, 
and the inferred stellar mass agrees with the observed luminosity.

%

\subsection{V645 Cyg}
The physical properties of the central star of V645~Cyg has been frequently
studied since its identification \citep{lebofsky76}.  It has a luminosity of
about $4\times10^4~L_{\sun}$ \citep{mottram}.  The near-infrared imaging
polarimetry shows a centro-symmetric pattern surrounding N0 nebula
\citep{minchin91}.  Some H$_2$O maser spots are detected exactly toward the N0
\citep{lada81}.  The spectral type of the central star was estimated to be O7e
\citep{cohen77}.  On the other hand, \cite{goodrich86} and \cite{hp89} detected
the P-Cygni profiles in \mbox{He\,{\sc i}}, \mbox{O\,{\sc i}}, \mbox{Si\,{\sc
    ii}}, and \mbox{Ca\,{\sc ii}}, suggesting Herbig Ae$/$Be stars.  Recently,
\cite{miroshnichenko09} assessed the effective temperature from the optical
spectroscopy.  They suggest that $T_\mathrm{eff}$ is about 25\,000~K, but that
the star is probably extended.  Given the variability in the spectrum, it is
not inconceivable that this result is in agreement with the earlier result of
\cite{clarke06}, who found a temperature nearer 30000K, and a type near O9Ve as
we do here (which is also more consistent with the known luminosity).  The lack
of HeII emission as clearly pointed out by \cite{miroshnichenko09} is not a
problem with this classification.

The P-Cygni profile is detected in Br$\gamma$ \citep[this work
  and][]{clarke06}.  The blueshifted absorption component ranges up to
$-800$~kms$^{-1}$ \citep[cf.\,$-2000$~kms$^{-1}$,][]{clarke06}.  A similar
profile is detected in W33A (D10).  Our spectroastrometry of Br$\gamma$ shows
no significant signal.  We suggest that this is due to the fact that Br$\gamma$
arises in a spherically outflowing wind in this source, which essentially has a
null spectroastrometric signature.  This also naturally explains the strong
P-Cygni profile.  We previously concluded that W33A has an asymmetric wind 
by comparison.  This suggests that V645~Cyg is actually nearer to its final
main sequence configuration.  Our fit to the CO bandhead also suggests that 
V645~Cyg must be close to pole-on.
%

\subsection{I19065 and G082}
These objects are identified as MYSOs by the RMS survey \citep{lumsden02} and
this work for the first time reports their individual study.  I19065 is a weak
HII region, with a small cluster of stars.  The brightest of those seen at $K$
may be the most important ionizing source.  In I19065, the \mbox{He\,{\sc i}}
line at 2.058~$\mu$m is detected in emission, as is Br$\gamma$.  The width of
the lines suggest an origin partly from the star as well as the HII region.
The estimated bolometric luminosity is 1.3$\times$10$^4~L_{\sun}$
\citep{mottram}.  We find good agreement between this luminosity and a possible
B1Ve classification from the properties of the HeI 2.058~$\mu$m line.
G082 has no special characteristic besides the detections of the Br$\gamma$
and \mbox{Fe\,{\sc ii}} emission lines in our data.  Taking into account
the estimated bolometric luminosity of 6.3$\times$10$^3~L_{\sun}$, this object
is probably a mid-B star.

\end{document}